\documentstyle[psfig,aaspp4,flushrt]{article}
\lefthead{Giovanelli et al.}
\righthead{No Hubble Bubble}
\def\be{\begin{equation}}
\def\ee{\end{equation}}
\def\etal{{\it et al. }}
\def\kms{km~s$^{-1}$~}

\def\dhh{$\Delta H_\circ/H_\circ$}

\def\W50{W$_{50}$~}
\received{Someday 1999}
\accepted{Someday 1999}
\journalid{337}{Someday 1999}
\articleid{11}{14}

\slugcomment{Submitted to {\it The Astrophysical Journal} in final form 8 May 1999}

\begin{document}
\title{No Hubble Bubble in the Local Universe}

\author {Riccardo Giovanelli, Daniel A. Dale, Martha P. Haynes}
\affil{Center for Radiophysics and Space Research
and National Astronomy and Ionosphere Center,
Cornell University, Ithaca, NY 14853}

\author {Eduardo Hardy}
\affil{National Radio Astronomy Observatory, Casilla 36--D, Santiago, Chile}

\author {Luis E. Campusano}
\affil{Observatorio de Cerro Cal\' an, Universidad de Chile, Casilla 36--D, Santiago, Chile}


\hsize 6.5 truein
\begin{abstract}
Zehavi et al. (1998) have suggested that the Hubble flow within 70$h^{-1}$ 
Mpc may be accelerated by the existence of a void centered on the Local Group. 
Its underdensity would be $\sim 20$\%, which would result in a local Hubble 
distortion of about 6.5\%. We have combined the peculiar velocity data of two 
samples of clusters of galaxies, SCI and SCII, to investigate the amplitude 
of Hubble distortions to 200$h^{-1}$ Mpc. Our results are not supportive of
that conclusion. The amplitude of a possible distortion in the Hubble flow 
within 70$h^{-1}$ Mpc in the SCI+SCII merged data is $0.010\pm0.022$.
The largest, and still quite marginal, geocentric deviation from smooth Hubble 
flow consistent with that data set is a shell with $\Delta H_\circ/H_\circ =0.027\pm0.023$, 
centered at $hd = 101$ Mpc and extending over some 30 $h^{-1}$ Mpc. Our results
are thus consistent with a Hubble flow that, on distances in excess of 
$\sim 50 h^{-1}$ Mpc, is remarkably smooth.

\end{abstract}



\section{Introduction}

The linearity of the Hubble law over large scales, as illustrated by the early
work of Sandage \& Hardy (1973), has been confirmed by more recent measurements
as discussed by Postman (1997). These measurements do not however exclude the
possibility of local deviations from Hubble flow with amplitudes on the order 
of a few percent, as would be produced by large--scale mass fluctuations. For
example, an extended, underdense region centered on the Local Group would exhibit 
a locally accelerated Hubble flow. This could, to a point, help reconcile 
discrepant estimates of the value of the Hubble constant obtained by methods 
which sample vastly different scales and solve the still raspy conflict between 
some estimates of the age of a matter dominated Universe and that of the oldest stars. By
analyzing the monopole of the peculiar velocity field as described by a sample
of 44 type Ia supernovae (SN), Zehavi \etal (1998, hereafter Z98) have recently 
suggested that the Local Group may be near the center of a bubble of 70$h^{-1}$ 
Mpc radius (where $H_\circ = 100 h$ \kms is the Hubble constant), underdense 
by 20\%, which may be itself surrounded by an overdense shell. The isotropic
flow observed within that ``bubble'' would then exceed the universal rate by 
\dhh$=(6.5\pm2.1)$\%; i.e., studies that rely on distance indicators contained
within that bubble would overestimate the Hubble constant by 6.5\% . Z98 
cautiously underscore the marginal character of their detection, as well as the 
need to corroborate, or refute, their suggestion by means of tests with independent 
sets of data. In this {\it Letter}, we provide such a test.

\section{The SCI and SCII Cluster Samples}

Based on data published earlier (Giovanelli \etal 1997a,b; hereafter G97a and 
G97b), peculiar velocities of 24 clusters of galaxies within 90$h^{-1}$ Mpc, 
obtained from measurements for 782 galaxies in their fields 
(hereafter referred to as SCI) have been recently presented by Giovanelli \etal 
(1998a), who also used it to compuete a dipole and investigate the Z98 claim. 
Because of the limited 
depth of the SCI sample, their test of the Z98 claim was inconclusive.
Recently, we have completed a deeper survey of cluster peculiar velocities, 
which extends to 200$h^{-1}$ Mpc. As for the SCI sample, the new survey is based 
on the Tully--Fisher (1977; hereafter TF) technique. The first installments 
of this data set are in Dale \etal (1997, 1998); the final one is in 
preparation, but in preliminary form its results can be seen in Dale (1998). 
The new survey, which we shall refer to as SCII, includes 522 galaxies in 
52 clusters. The dipole signature of the SCII, which is consistent with that of 
the CMB temperature dipole, is discussed in Dale \etal (1999).

The combination of SCI and SCII provides a peculiar velocity data set of
slightly smaller depth, but higher sampling density than the SN sample
of Z98. The peculiar velocity errors of the SCI set vary between 3\% and 6\%
of the distance, for each individual cluster. In the case of the SCII set,
peculiar velocity errors are somewhat higher, due to the smaller number of
galaxies observed per cluster: they hover between 4\% and 9\%, except
in a few cases which will be discussed later. On the average, the accuracy
of each cluster peculiar velocity compares favorably with the quoted uncertainty
of 5--8\% (for the internal errors alone) of the distance of individual SN 
in the Z98 sample. Since the 76 clusters in the SCI+SCII merged sample straddle 
quite comfortably the boundaries of the Z98 bubble, they can provide tighter 
constraints than the SN sample on the amplitude of the proposed, locally 
underdense region.

Table 1 lists the clusters in the SCI+SCII merged sample, identified either
by their Abell number (Abell, Corwin \& Olowin 1989) or by their common name,
the adopted center coordinates, as well as the radial velocity $cz_{cmb}$ and 
the peculiar velocity $V_{pec}$ in the 
CMB reference frame, (after Giovanelli \etal 1998a and Dale 1998) and the number 
$N$ of galaxies in each cluster with TF measurements. We compute 
a distance  $h d = (cz_{cmb} - V_{pec})/100$ and 
a deviation from Hubble flow
$\Delta H_\circ/H_\circ = V_{pec}/(cz_{cmb} - V_{pec})$. 

The total error on the peculiar velocity of each cluster, as listed in Table 1, 
includes several components, arising from: (i) photometric and spectroscopic
observational errors; (ii) uncertainties in the corrections applied to observed
parameters; (iii) uncertainties in the cluster redshifts; (iv) the scatter in the 
TF relation; (v) uncertainties in the TF template relation slope and zero point,
especially that deriving from the assumed standard of rest. We discuss point (v)
in greater detail in the next Section. The other sources of error are 
extensively discussed in the data papers mentioned above.

\section{Template Relation Accuracy and its Effect on the Monopole Moment}      

TF peculiar velocities are derived as offsets from a template relation,
which in its simplest form is defined by two parameters: a slope and a zero 
point. Errors on both the zero point and on the slope translate into spurious, 
geocentric peculiar velocity fields. For example, an error of 0.05 mag in the 
zero point would simulate a slowdown or speeding up of the Hubble expansion 
by 2.3\%. As for the effect of an error on the TF slope: if the template relation 
is, for example, too steep --- i.e. for a given velocity width which is broader 
than some fiducial value the template predicts too bright a magnitude ---, then 
high width galaxies will preferentially yield positive magnitude offsets. The
opposite will be true for low width galaxies. 
Since low width galaxies are intrinsically faint, they 
are more likely to be present in nearby samples than in more distant ones; thus 
nearby samples fitted with too steep a TF template relation exhibit a net negative
magnitude offset, which translates into a spurious outflow. The effect of unrecognized 
TF calibration errors can then be misconstrued as a monopole perturbation, and thus 
as a geocentric Hubble flow distortion.

The TF template relation is determined {\it internally} for a cluster sample.
In the case of SCI, it was obtained by assuming that the subset of clusters
farther than 40$h^{-1}$ Mpc has a globally null monopole (G97b). Dale (1998)
obtained an SCII template by assuming that the set of clusters has a
globally null monopole, and adopting the same TF slope as for the SCI sample.
As discussed in G97b, given a number $N$ of clusters the uncertainty on the 
TF zero point of the resulting template cannot be depressed indefinitely by 
increasing the average number $\bar n$ of galaxies observed per cluster, and 
taking advantage of the $\bar n^{-1/2}$ statistical reduction of noise on the 
mean. That is because a ``kinematical'' or ``thermal''  component of the 
uncertainty depends on the number 
$N$, distribution in the sky and peculiar velocity distribution function of the 
clusters used. In SCI, for example, the statistical uncertainty deriving from 
the total number of galaxies observed ($\bar n\times N$) is exceeded by the kinematic
uncertainty, which is quantified as follows. For a sample of $N$ clusters of 
average redshift $<cz>$, the most probable systematic error on the template 
relation zero point is 
$|\Delta m| \simeq 2.17 <V_{pec}^2>^{1/2} <cz>^{-1} N^{-1/2}$, where 
$<V_{pec}^2>^{1/2}$ (expressed in the same units as $cz$) is the line of sight
r.m.s. cluster peculiar velocity, of about 300 \kms ~(G97b; Giovanelli \etal 1998a; 
Dale 1998). This quantity is about 0.04 mag for SCI, while it is only 0.01 mag 
for SCII 
due to the larger mean distance and number of clusters of the latter. Since 
the total number of galaxies involved in the two samples is comparable,
the zero point of the SCII template is thus more accurate than that of SCI.
On the other hand, the peculiar velocities of individual clusters in SCII are
less accurate than those in SCI. We note that the kinematical or thermal 
component of the uncertainty is larger for SN peculiar velocities than for
our cluster ones. That is because the amplitude of the distribution function
of peculiar velocities among individual galaxies --- the hosts of SN --- is
larger than that of clusters, as the former is amplified by the variance
associated with fluctuations on small scales.

In the case of both SCI and SCII, a direct TF template relation was obtained, using
the approach described in G97b. The data for each cluster offset was corrected for 
the effect of an incompleteness bias. The zero points of the two templates were 
found to agree to within 0.015 mag (SCII being fainter by that amount).

In this paper, we combine the SCI and SCII samples, and use them to investigate
the presence of large scale variations in the monopole of the Hubble flow. Note
that such combined sample cannot be used for the detection of a geocentric
deviation from smooth Hubble flow{\it  which would extend over the full volume 
sampled by the total cluster set, as it would be null by design}.
The merged cluster data set can however be used to detect {\it changes in} 
$\Delta H_\circ/H_\circ$ that would take place well within the volume spanned 
by the data. The amplitude of the change (say a step in $\Delta H_\circ/H_\circ$) 
that can be detected depends on the location of the presumed step and on the 
accuracy with which the match in the TF zero point between the SCI and SCII samples
is established: 

\noindent
(i) For our cluster sample, a step would be ideally situated between 70$h^{-1}$ 
and 110$h^{-1}$ Mpc, in order to maximize the chance of detection, because it 
would split the cluster sample into two roughly equal parts. The SCI+SCII sample 
is thus well-suited to test the Z98 result.

\noindent
(ii) The internal accuracy of the zero point for the SCI sample is 0.025 mag;
however, since it is based on a subset of 14 clusters farther than 40$h^{-1}$ 
Mpc, the kinematical uncertainty of 0.04 mag, as mentioned above, increases the
total uncertainty to 0.045 mag. The total uncertainty on the zero point of SCII, 
because it involves a larger number of more distant clusters, is only 0.025 mag; 
the kinematical component in this case is only 0.01 mag. Note for comparison 
that a 6.5\% step in $\Delta H_\circ/H_\circ$ would translate in a 0.13 mag 
differential TF offset between clusters on each side of the step.

It is also useful to point out that each of the two samples was completed over 
many observing runs, both in their photometric and spectroscopic parts, and
a number of objects were observed in more than one run. Mismatches in the
cross--run and cross--cluster calibrations thus have been minimized and their 
impact on the final error budget is included in the statistical estimate
given above. 

\noindent
(iii) The small overlap in distance between SCI and SCII occurs near 70$h^{-1}$ 
Mpc, which is the edge of the Hubble bubble suggested by Z98 (4 clusters in
SCI are farther than $cz=7000$ \kms, while 4 in SCII are within that redshift).
We thus need to establish the impact of the accuracy of the match between the 
two samples' zero points, on the estimation of the likelihood of a Hubble bubble. 
We return to this point in Section 4.1.

\section{Geocentric Hubble Deviations}                               

Using the data in Table 1, and forcing the template TF zero point to be the 
same for SCI and SCII, we obtain Figure 1, a plot of the Hubble deviation
versus the distance. In the upper panel of Figure 1 we display the individual
data points, while in the lower one we show the errors associated with each
measurement. Starred symbols refer to the SCI sample, while circular symbols
refer to SCII. Eight clusters, flagged by double daggers in Table 1, are plotted 
in Figure 1 as unfilled symbols: their peculiar velocities have been obtained 
from fewer than five TF measurements and are thus very unreliable. The latter 
are not used in the following statistical analyses.

The plot presented in Figure 1 is similar to that in Figure 1 of Z98. For
comparison, we have included the outline of the Z98 step as a dashed line, 
which extends from zero to 70$h^{-1}$ Mpc distance, at the level of 
$\Delta H_\circ/H_\circ = 0.065$. We note immediately that the Z98 proposal
of a Hubble bubble is not corroborated by the cluster data. We also note that
at distances nearer than $\sim 30 h^{-1}$ Mpc even modest peculiar velocities 
constitute a sizable fraction of $cz$, thus amplifying and distorting the values 
of $\Delta H_\circ/H_\circ$. The implied deviation from Hubble flow they reveal 
is of scarce interest, as they apply to too small, too sparsely sampled a volume.

Next, we test for the presence of a step at 70$h^{-1}$ Mpc distance, of
the kind suggested by Z98, and  we search for the signature of other possible,
geocentric large--scale fluctuations in the Hubble flow.

\subsection{Test for a Hubble Bubble}                               

We consider whether a step is present in $\Delta H_\circ/H_\circ$ at 70$h^{-1}$ Mpc,
by taking the difference in the average of $\Delta H_\circ/H_\circ$ between 
$30h^{-1}$ and $70h^{-1}$ Mpc, and the corresponding average at distances higher than  
$70h^{-1}$ Mpc. That difference is $0.010\pm0.012$, if individual clusters are
weighed by their errors in  $\Delta H_\circ/H_\circ$, and $0.007\pm0.012$ if
equal weight averages are computed. The uncertainty of this result can however
be affected by a number of systematic errrors, which exceed the statistical
estimate given above; in the following we discuss them one by one.

{\it Kinematic Zero Point Mismatch:} 
First, we consider the impact of the systematic mismatch between the TF template 
zero points of
the two samples, as discussed in part (iii) of Section 3. We can evaluate the
impact of that uncertainty on the determination of the amplitude of a possible 
step at $70h^{-1}$ Mpc by offsetting by $\pm(0.04^2+0.01^2)^{1/2}= \pm0.04$ 
mag the SCI and SCII samples with respect to each other, computing in each case 
the amplitude of the step (note that there are SCI and SCII clusters on both sides
of the step). The results are respectively $0.022$ and $-0.004$ mag. It can thus
be inferred that the impact on the uncertainty of the step, produced by a possible 
systematic error in the match between zero points for the two samples, is about
0.03 mag or 1.5\%.

{\it Differential Malmquist Bias:}
Malmquist bias corrections have not been applied to the cluster peculiar velocities.
If such a correction were the same for all the clusters, it would have no impact
on the detectability of a Hubble bubble step. However, since the more distant
clusters of SCII each include a smaller number of galaxies with TF measurements,
the impact of a possible differential Malmquist bias between SCI and SCII needs
to be explored. As discussed in Giovanelli \etal (1998a), the Malmquist bias can be 
estimated with adequate accuracy in the "homogeneous" assumption, i.e. that the 
clusters' distribution in space is Poissonian, and shown to be quite small. The
Malmquist bias correction in that case is $e^{3.5\Delta^2}-1$, where 
$\Delta=10^{0.2\epsilon/\sqrt(n)} - 1$, with $\epsilon$ the scatter in magnitudes
about the TF relation (about 0.35 mag) and $n$ the number of galaxies with TF
measurements per cluster. For 
example, for a cluster with 10 galaxies with TF measurements, the average for SCII, 
the Malmquist bias correction is 1.0\% on the distance. In the case of SCI, the
average number of galaxies with TF measurements per cluster is about 16. In that
case the Malmquist bias correction is 0.7\% on the distance. Neglect to apply
a Malmquist bias correction thus introduces a possible bias with an amplitude
of 0.003 in $\Delta H_\circ/H_\circ$. 

{\it Template Relation Slope:}
The same template relation slope has been used for both SCI and SCII, as 
discussed in Dale (1998). The error on the determination of that slope is
given in G97b, as 0.12 on a slope of -7.68, or 1.6\%. If there were a sgnificant
difference in the distribution
of galaxies as a function of velocity width, between nearby and more distant 
clusters, the uncertainty on the slope would introduce a systematic
bias in the distances. To estimate the amplitude of that bias, we 
binned galaxies as a function of width, separately for 
the clusters within and beyond 70$h^{-1}$ Mpc, and for each group estimated
the average magnitude offset introduced by an error in the slope of 1.6\%;
in doing so, we assumed that the zero point, i.e. the value of the template relation
at $\log W = 2.5$, is correct. The resulting TF offset uncertainty between the 
two groups is 0.0055 mag, or 0.0025 on the distance. 

{\it Evolution:}
Some authors (Rix \etal 1997; Simard \& Pritchet 1998) have claimed substantial
evolution in the mass--to--light ratio of spiral galaxies between $z=0$ and
relatively modest redshifts $s\sim 0.4$, while others (Vogt \etal 1997; Bershady
1996; Dale, Us\' on \& Giovanelli 1999, in preparation) find no such effect. 
Evolution would 
translate into a shift of the TF relation zero point. While this issue is still 
quite uncertain, we can estimate the possible impact of evolution, assuming a 
(rather generous) shift of 1 mag between $z=1$ and $z=0$. The difference
in $z$ between the clusters within 70$h^{-1}$ Mpc and those farther away is
$\simeq 0.02$, thus a possible shift of 0.02 mag or 0.01 in distance would be
possible. The direction of this relative shift would be that of a gradual 
brightening of the higher redshift galaxies and therefore increasing their 
average $\Delta H_\circ/H_\circ$. Probably overestimating it, we conclude that 
the uncertainty associated with this effect is 0.01 in the distance.

Below, we give a summary of the components of uncertainty with which the SCI+SCII
merged sample can be used to identify a possible step in the Hubble flow at 
70$h^{-1}$ Mpc:

\vskip 0.3in
\begin{tabular}{llrr} \hline
& statistical distance error & 1.2\% & \\

& kinematic zero point mismatch between SCI and SCII & 1.5\% & \\

& Malmquist bias  & 0.3\%  & \\

& Template slope uncertainty & 0.3\% &  \\

& Evolution  & 1.0\%  & \\ \hline

& Total &  2.2\% & 
\end{tabular}

Thus our estimate of the amplitude and significance of a step in the Hubble
flow at 70$h^{-1}$ Mpc is
\be
{\Delta H_\circ \over H_\circ} = 0.010\pm0.022 \nonumber
\ee

For a two--zone model, which includes an inner void out to 70$^{-1}$ Mpc and an 
outer region expanding at the Hubble rate, Z98 report an amplitude of $0.065\pm 
0.021$; such a void is not apparent in our data, in which a 6.5\% step would be a  
3--$\sigma$ event. We remark however that, compounding our estimated error with 
that reported in Z98, the difference between the two results is $0.055\pm0.030$.

The Hubble distortion 
reported by Z98 reduces to $0.053\pm0.022$ in a three--zone model, where an inner 
underdense sphere of 70$h^{-1}$ Mpc is surrounded by an overdense shell between 
70 and 105 $h^{-1}$ Mpc; for the outer shell, Z98 report an inflow of $\Delta 
H_\circ/H_\circ = -0.059\pm0.027$. For the latter region, between 70 and 105 
$h^{-1}$ Mpc, our data yields $\Delta H_\circ/H_\circ = +0.020\pm0.018$. The
difference between our and the Z98 results, compounding the reported errors,
is $0.079\pm0.032$.

\subsection{Other Geocentric Deviations}

We search for the signature of other possible large--scale fluctuations
in the Hubble flow, adopting a similar $\chi^2$ minimization analysis to
that carried out by Z98. The minimization of 
\be
\chi^2 = \sum_i \Bigl[\log [1+(\Delta H_\circ/H_\circ)_i] - \log [1+(\Delta H_\circ/ H_\circ)_{model}] 
\Bigr]^2/\sigma_i ,
\ee
where $(\Delta H_\circ/H_\circ)_i$ are the values plotted in Figure 1 and
$(\Delta H_\circ/ H_\circ)_{model}$ is a model with a constant departure from
zero in $\Delta H_\circ/ H_\circ$ between two arbitrary distances $hd_1$ and
$hd_2$, is equivalent to that expressed in Eqn. (2) of Z98. $\sigma_i$ is
the estimated error on $\log [1+(\Delta H_\circ/ H_\circ)_i]$. The strongest
signature for a departure from Hubble flow consistent with the SCI+SCII cluster 
data is 
a (very marginal) step of amplitude $\Delta H_\circ/H_\circ =0.027\pm0.023$ centered 
at $hd = 101$ Mpc and 33 Mpc wide. The boundaries of the region are very ``soft''. 
In the calculations, we impose that the width of the perturbed region
should be no less than 20 Mpc, including only 61 clusters with $h d$ between 
30 and 200 Mpc and excluding the eight clusters with poor sampling ($n<5$) 
plotted as unfilled symbols in Figure 1 and flagged in Table 1.

\begin{figure}
\plotone{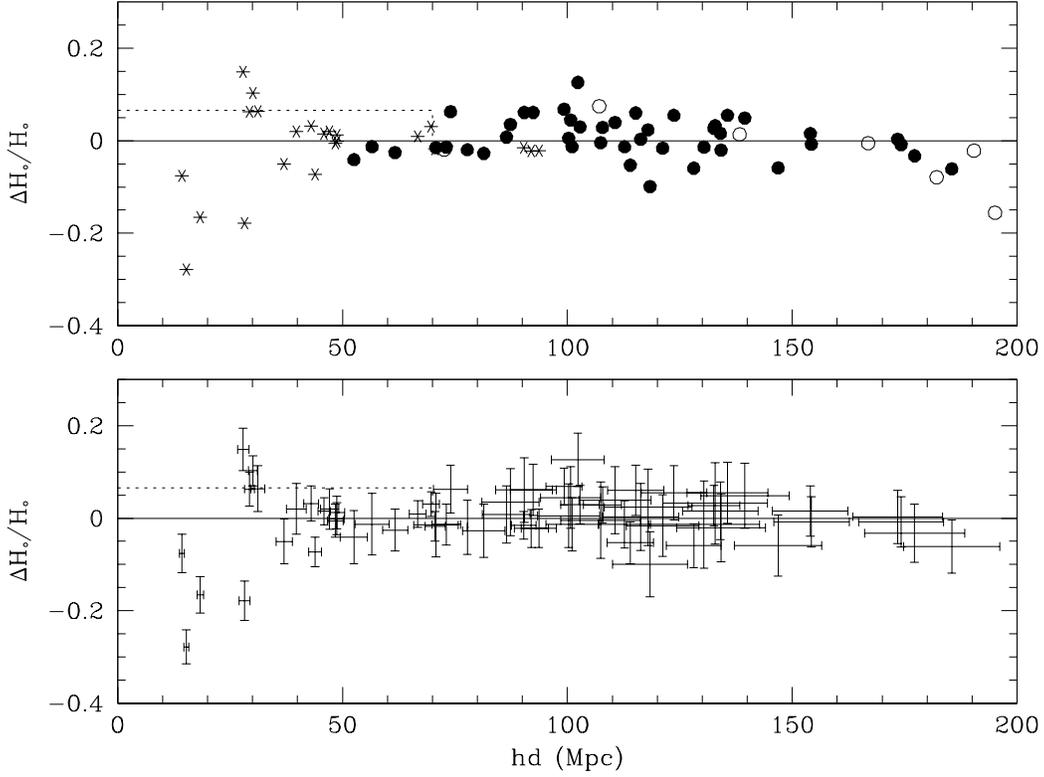}
\caption{Deviations from Hubble flow plotted versus TF distance
for the clusters listed in Table 1. In the upper panel, starred symbols
refer to clusters in SCI, while circular symbols refer to SCII. Among the 
latter, filled symbols
identify clusters with distance determinations based on $n>4$ individual
galaxy TF distances, while unfilled ones refer to clusters with $n\leq 4$, 
the peculiar velocities of which are deemed least trustworthy and are not 
used in the statistical analysis; their names are flagged in Table 1. 
The horizontal dashed line identifies the acceleration of 6.5\% in the Hubble 
flow within $hd=70$ Mpc claimed by Z98. The lower panel displays the error
bars associated with each starred or filled data point.}
\end{figure}

\section {Conclusions}

Z98 have cogently argued that a region of $70^{-1}$ Mpc radius could be
underdense by $\sim 20$\%  --- which is the amount necessary to produce a 
suggested local acceleration of 6.5\% of the Hubble flow ---, without 
unreasonably stretching the plausible amplitude range of 
cosmological density fluctuations.
One would be left, of course, with the nagging coincidence of the nearly central 
location of the LG in the void (a circumstance which would also be at some 
odds with the fairly large peculiar velocity of the LG of some 620 \kms, as 
indicated by the CMB dipole). 
Our data give an amplitude for a possible Hubble flow distortion within 
70 $h^{-1}$ Mpc of $\Delta H_\circ / H_\circ = 0.010\pm0.022$. 

In a three--zone model, Z98 suggest that an overdense shell between $70^{-1}$ 
and $105h^{-1}$ Mpc may be affected by an inflow of $\Delta H_\circ/H_\circ = 
-0.059\pm0.027$. For that region, our data yields $\Delta H_\circ/H_\circ = 0.020\pm0.018$.

The distortion of largest amplitude, consistent 
with our data, is $\Delta H_\circ/H_\circ =0.027\pm0.023$ centered at $hd = 101$ 
Mpc and extending over a shell some 30 $h^{-1}$ Mpc thick. 

The results of this paper are consistent with those on the peculiar velocity 
field as traced by the SFI sample of field spirals: its dipole converges to
that of the CMB dipole, both in amplitude and apex direction, within about
$50 h^{-1}$ Mpc (Giovanelli \etal 1998b).
We conclude that, at distances in excess of $\sim 50 h^{-1}$ Mpc, the cluster 
peculiar velocity data are consistent with a picture in which the
average Hubble flow is remarkably smooth.

\acknowledgements

The comments of an anonymous referee helped improve the presentation of this paper.
The results presented in this paper are based on observations carried out at
the Arecibo Observatory, which is part of the National Astronomy and 
Ionosphere Center (NAIC), at the Kitt Peak National Observatory (KPNO), the 
Cerro Tololo Interamerican Observatory (CTIO) and at the Palomar Observatory (PO), 
NAIC is operated by Cornell University, KPNO and CTIO by Associated Universities 
for Research in Astronomy, all under cooperative agreements with the National 
Science Foundation. The Hale telescope at the PO is operated by 
the California Institute of Technology under a cooperative agreement with 
Cornell University and the Jet Propulsion Laboratory.  
This research was supported by NSF grants AST94--20505 and AST96--17069 to RG, 
AST95-28860 to MH and by Fondecyt grant 1970735 to LEC.

\newpage

\begin{deluxetable}{lrrrrrr}
\tablewidth{0pt}
\scriptsize
\tablenum{1}
\tablecaption{Cluster Positions and Velocities}
\tablehead{
\colhead{Cluster}   & RA(1950) & Dec (1950) & \colhead{$cz_{cmb}$}  &  
\colhead{$V_{pec}$} &
\colhead{$n$}       
}
\startdata
{\bf SCI}   &           &         &        &                    &     \nl
\\
N383        &  010430.0 &   +321200 &  4865 $\pm$ 32  &    -6 $\pm$   170  &  21 \nl
N507        &  012000.0 &   +330400 &  4808 $\pm$ 99  &    94 $\pm$   204  &  14 \nl
A262        &  014950.0 &   +355440 &  4664 $\pm$ 80  &    70 $\pm$   133  &  31 \nl
A400        &  025500.0 &   +055000 &  6934 $\pm$ 75  &  -126 $\pm$   227  &  25 \nl
Eridanus    &  033000.0 & $-$213000 &  1534 $\pm$ 30  &  -304 $\pm$    74  &  34 \nl
Fornax      &  033634.0 & $-$353642 &  1321 $\pm$ 45  &  -109 $\pm$    60  &  39 \nl
Cancer      &  081730.0 &   +211400 &  4939 $\pm$ 80  &    61 $\pm$   172  &  26 \nl
Antlia      &  102745.0 & $-$350411 &  3120 $\pm$100  &   185 $\pm$   109  &  27 \nl
Hydra       &  103427.7 & $-$271626 &  4075 $\pm$ 50  &  -320 $\pm$   142  &  25 \nl
N3557       &  110735.0 & $-$371600 &  3318 $\pm$ 57  &   199 $\pm$   155  &  11 \nl
A1367       &  114154.0 &   +200700 &  6735 $\pm$ 88  &    62 $\pm$   191  &  35 \nl
Ursa Major  &  115400.0 &   +485300 &  1101 $\pm$ 40  &  -425 $\pm$    56  &  30 \nl
Cen30       &  124606.0 & $-$410200 &  3322 $\pm$150  &   310 $\pm$    98  &  38 \nl
A1656       &  125724.0 &   +281500 &  7185 $\pm$ 68  &   212 $\pm$   210  &  41 \nl
ESO508      &  130954.0 & $-$230854 &  3210 $\pm$100  &   417 $\pm$   128  &  17 \nl
A3574       &  134606.0 & $-$300900 &  4817 $\pm$ 30  &   -26 $\pm$   174  &  20 \nl
A2197\dag   &  162630.0 &   +410100 &  9162 $\pm$100  &  -204 $\pm$   384  &  25 \nl
Pavo II     &  184200.0 & $-$632000 &  4444 $\pm$ 70  &   137 $\pm$   163  &  18 \nl
Pavo        &  201300.0 & $-$710000 &  4055 $\pm$100  &    80 $\pm$   219  &  10 \nl
MDL59       &  220018.0 & $-$321400 &  2317 $\pm$ 75  &  -503 $\pm$   120  &  23 \nl
Pegasus     &  231742.6 &   +075557 &  3519 $\pm$ 80  &  -186 $\pm$   180  &  17 \nl
A2634       &  233554.9 &   +264419 &  8895 $\pm$ 79  &  -136 $\pm$   270  &  26 \nl
A2666       &  234824.0 &   +264824 &  7776 $\pm$ 84  &  -156 $\pm$   459  &   9 \nl
\\
{\bf SCII}  &           &         &        &                    &     \nl
\\
A2806       &  003754 & $-$562600 &  7867 $\pm$ 80  &   464 $\pm$   382  &  10 \nl
A114        &  005112 & $-$215800 & 17144 $\pm$143  &  -578 $\pm$  1111  &   9 \nl
A119        &  005348 & $-$013200 & 13141 $\pm$ 85  &  -275 $\pm$   988  &   6 \nl
A2877       &  010736 & $-$461000 &  6974 $\pm$ 58  &  -104 $\pm$   489  &   7 \nl
A2877b      &  010736 & $-$461000 &  9040 $\pm$ 48  &   307 $\pm$   634  &   5 \nl
A160        &  011012 & $+$151500 & 12072 $\pm$141  &   280 $\pm$   977  &   6 \nl
A168        &  011236 & $-$000100 & 13049 $\pm$ 58  &   679 $\pm$   725  &   9 \nl
A194        &  012300 & $-$014600 &  5037 $\pm$ 37  &  -216 $\pm$   302  &  13 \nl
A260        &  014900 & $+$325500 & 10664 $\pm$111  & -1175 $\pm$   835  &   9 \nl
A397        &  025412 & $+$154500 &  9594 $\pm$ 78  &   553 $\pm$   630  &  14 \nl
A3193       &  035654 & $-$522900 & 10522 $\pm$112  &   450 $\pm$   668  &   6 \nl
A3266\ddag  &  043030 & $-$613500 & 17782 $\pm$ 61  & -2700 $\pm$  2345  &   2 \nl
A496        &  043118 & $-$132100 &  9809 $\pm$ 59  &   566 $\pm$   513  &   9 \nl
A3381\ddag  &  060806 & $-$333500 & 11510 $\pm$ 48  &   798 $\pm$   868  &   4 \nl
A3407       &  070342 & $-$490000 & 12861 $\pm$136  &  -179 $\pm$  1235  &   8 \nl
A569        &  070524 & $+$484200 &  6011 $\pm$ 43  &  -157 $\pm$   280  &  13 \nl
A634        &  081030 & $+$581200 &  7922 $\pm$ 42  &  -222 $\pm$   469  &   8 \nl
A671        &  082524 & $+$303500 & 15307 $\pm$194  &  -120 $\pm$   838  &   9 \nl
A754\ddag   &  090624 & $-$092600 & 16599 $\pm$ 82  &   -92 $\pm$  3294  &   3 \nl
A779        &  091648 & $+$335900 &  7211 $\pm$101  &  -100 $\pm$   320  &  14 \nl
A957        &  101124 & $-$004000 & 13819 $\pm$120  &  -866 $\pm$   974  &   6 \nl
A1139       &  105530 & $+$014600 & 12216 $\pm$ 71  &   694 $\pm$   629  &  11 \nl
A1177       &  110648 & $+$215800 & 10079 $\pm$ 81  &    51 $\pm$   689  &   6 \nl
A1213       &  111348 & $+$293200 & 14304 $\pm$ 90  &   744 $\pm$   899  &   6 \nl
A1228       &  111848 & $+$343600 & 10794 $\pm$ 34  &  -603 $\pm$   517  &  13 \nl
A1314       &  113206 & $+$491900 &  9970 $\pm$154  &  -134 $\pm$   582  &   8 \nl
A3528\ddag  &  125136 & $-$284500 & 16770 $\pm$139  & -1441 $\pm$  1703  &   3 \nl
A1736       &  132406 & $-$265100 & 10690 $\pm$ 50  &   -49 $\pm$   887  &   6 \nl
A1736b\ddag &  132406 & $-$265100 & 14017 $\pm$ 84  &   186 $\pm$  1121  &   4 \nl
A3558       &  132506 & $-$311400 & 14626 $\pm$ 44  &   678 $\pm$   981  &   8 \nl
A3566       &  133606 & $-$351800 & 15636 $\pm$ 87  &   236 $\pm$   837  &   9 \nl
A3581\ddag  &  140436 & $-$264700 &  7122 $\pm$126  &  -139 $\pm$   659  &   4 \nl
A1983b      &  144724 & $+$170600 & 11524 $\pm$ 62  &  1291 $\pm$   589  &   8 \nl
A1983       &  145024 & $+$165700 & 13715 $\pm$ 45  &   429 $\pm$  1165  &   7 \nl
A2022       &  150212 & $+$283700 & 17412 $\pm$ 72  & -1134 $\pm$  1067  &   8 \nl
A2040       &  151018 & $+$073700 & 13616 $\pm$ 61  &   212 $\pm$   839  &  10 \nl
A2063       &  152036 & $+$084900 & 10605 $\pm$ 53  &   680 $\pm$   398  &  18 \nl
A2147       &  160000 & $+$160200 & 10588 $\pm$ 85  &   303 $\pm$   427  &  19 \nl
A2151       &  160300 & $+$175300 & 11093 $\pm$ 59  &   312 $\pm$   424  &  22 \nl
A2256       &  170636 & $+$784700 & 17401 $\pm$132  &    56 $\pm$   998  &   8 \nl
A2295b\ddag &  175900 & $+$691600 & 18633 $\pm$ 82  &  -408 $\pm$  1587  &   4 \nl
A2295       &  180018 & $+$691300 & 24554 $\pm$199  & -1145 $\pm$  1448  &  10 \nl
A3656       &  195712 & $-$384000 &  5586 $\pm$ 64  &   -72 $\pm$   375  &   6 \nl
A3667\ddag  &  200830 & $-$565800 & 16477 $\pm$ 94  & -3034 $\pm$  1582  &   4 \nl
A3716       &  204754 & $-$525400 & 13618 $\pm$ 64  &   359 $\pm$   581  &  14 \nl
A3744       &  210418 & $-$254100 & 11123 $\pm$ 89  &  -150 $\pm$   578  &  11 \nl
A2457       &  223312 & $+$011300 & 17280 $\pm$110  &  -144 $\pm$   946  &   9 \nl
A2572       &  231554 & $+$182800 & 11495 $\pm$100  &   436 $\pm$   803  &   5 \nl
A2589       &  232130 & $+$163300 & 11925 $\pm$ 95  &  -194 $\pm$   804  &   6 \nl
A2593       &  232200 & $+$142200 & 12049 $\pm$ 86  &  -761 $\pm$   605  &  12 \nl
A2657       &  234218 & $+$085200 & 11662 $\pm$137  &    32 $\pm$   844  &   5 \nl
A4038       &  234506 & $-$282500 &  8713 $\pm$ 63  &    68 $\pm$   534  &   7 \nl
\enddata
\tablenotetext{\dag} {Includes A2197, A2199.}
\tablenotetext{\ddag}{Excluded from statistical analysis.}
\end{deluxetable}

\vfill
\end{document}